# Attaining High Bandwidth In Cloud Computing Through SDN-enabled Multi-tree Multicast

Sayantan Guha[#1], and Adel Alshamrani[*2]

[#]CIDSE, Arizona State University, Tempe, USA
[*]CCSE, University of Jeddah, Jeddah, Saudi Arabia

***Abstract*** — *Achieving high bandwidth utilization in cloud computing is essential for better network performance. However, it is difficult to attain high bandwidth utilization in cloud computing due to the complex and distributed natures of cloud computing resources. Recently, a growing demand for multicast transmission is perceived in cloud computing, due to the explosive growth of multi-point communication applications, such as video conferencing, online gaming, etc. Nonetheless, the inherent complexity in multicast routing in cloud computing, existing multicast plans failed to produce effective and efficient protocol schemes, which limits the application of multicast communication on the Internet. In this paper, a technique is proposed in how the newly developed network architecture, Software Defined Network (SDN), can promote the design of the multicast protocol and improve the performance of the multicast transmission in the cloud computing. The approach is to use the SDN-cloud Computing-enabled multicast communication scheme with ultra-high bandwidth utilization. The bandwidth utilization is enhanced by measuring various routing trees for each multicast transmission session and distributing the traffic load over all available routes in the cloud computing resources. The SDN is utilized to tackle with various design hurdles in the cloud computing, including both the current ones with the conventional multicast pattern and the newly emerged ones with multi-tree multicast. The prototype implementation and experiments demonstrate the performance enhancement of the proposed approach in the cloud computing in compared to conventional single-tree multicast designs.*

## I. INTRODUCTION

With the advent of cloud computing, the efficiency to process the data at the cloud of the network system has increased as more data is generated at the cloud of the network. Since the birth of the Internet, unicast communication has governed the Internet. However, lately, there has been an intensifying desire for multicast transmission due to the evolution of applications with multi-point communication specifications, such as video conferencing, parallel computing, online gaming, etc. These multi-point communications also impact the cloud computing in terms of bandwidth availability of the network systems.

However, due to the intrinsic complexity of multicast routing and transportation, few of the existing multicast schemes remain in practice in the current cloud computing. This ineffectiveness and inefficiency of existing multicast designs in cloud computing exceedingly limit the application of multicast transportation in the current cloud computing architecture. There are many factors contributing to the ineffectiveness and the low performance of traditional multicast schemes. For instance, the distributed nature of the cloud computing architecture restricts the scope of each entity in the cloud computing in making routing/forwarding decisions. Each router in the cloud computing results routing algorithms based on information transferred from nearby routers in cloud computing, doing it inefficient in dealing with network dynamics. Similarly, with only the local view of the cloud computing, each switch is not able to make global optimal forwarding choices regarding the network circumstances. While the inefficiency brought about by the limited view is still tolerable in unicast communication, it is definitely non-negligible in multicast due to the added dimension in the problem. Another notable concern is the scalability.

Multicast is disreputable for its poor scalability brought by the necessity in performing dynamic group management. In most existing multicast schemes for cloud computing, the group management functionality is performed by the routing or forwarding elements in the cloud network. The routers/switches monitor and process the cloud network for group joining and leaving messages, calculate the optimal routes/forwarding decisions, and perform the routing/forwarding functionality based on the cloud network view of their local scopes. This causes a prominent burden to the routing/forwarding components in the cloud network. Other contributors to multicast's low performance are: the natural complexity in computing multicast routes, the lack of efficient reliable protocol [1], [2], the hardness in satisfying multiple QoS constraints [3], etc. While these problems are hard to solve, in this paper the previous two practical problems of multicast communication in cloud computing are focused. In this paper, a multi-tree multicast communication scheme is proposed with high bandwidth utilization. The proposed technique aims to utilize the path diversity in modern network topologies for cloud computing. For each multicast session, multiple multicast trees are calculated, each with a specific bandwidth share on the tree. The sender distributes the total traffic over all the multicast trees that it can utilize, thus achieving much higher bandwidth utilization than the conventional single-tree schemes.

There are several challenges in designing a multi-tree multicast scheme in cloud computing. First, assigning multiple trees to a single multicast session can further deteriorate the scalability of multicast in cloud network. While previously each switch/router need only to store information about the single-tree route, now multiple routes need to be stored, and also more switches may be involved in the transmission of one multicast session. Second, computing multiple multicast trees requires more computation resources and wider view of the cloud network. While computing a single routing tree can be performed in a distributed manner, it is hard for the switch/router to calculate many trees at the same time with its limited computation resource and scope of view.

Theoretically, computing multiple cloud-disjoint Steiner Trees is algorithmically harder than computing a single multicast tree [4]. Third, distributing the traffic load over multiple trees in cloud networks needs knowledge about the properties of the trees on the sender side, such as bandwidth, delay etc. While in single-tree, routing information gathering is relatively trivial as there are many works have been done [5], [6] and [7]. It is unknown whether gathering information of multiple trees is practical when the gathering is done in a distributed manner. Unevenly, distributing the traffic may lead to hot-spots in the network, degrading the benefit of multi-tree multicast. Fourth, multi-tree multicast may cause the problem of packet out-of-order delivery at the receiver side. This is due to the fact that in single-tree schemes the transmission of packets are in a sequential manner, while in multi-tree multicast the transmission of multiple packets occurs in a parallel manner. This phenomenon is further exacerbated when the multiple trees have various latency properties. This may cause great problems in some real-time applications in cloud network such as video conferencing etc.

To tackle these challenges, Software Defined Networking (SDN) [8], a newly emerged network control plane architecture, is exploited in the proposed system design in cloud computing. SDN has the natural advantage in dealing with challenges above for cloud network. With the flexible control offered by SDN, the proposed architecture can offload most of the computation and management tasks to the controller, leaving only the forwarding functionality in network intermediate nodes, thus solving both the scalability problem and the computational limitations. For the fourth challenge, a well-designed transportation the protocol can solve the problem of reliability, which is not within the scope of this paper. In general, with the help of SDN, it is applicable to design a much more effective and efficient multicast scheme compared to the traditional schemes for cloud networks.

The paper is organized as follows. Section II gives an overview of the proposed design for cloud network. Section III demonstrates our method in designing the proposed scheme. Section IV presents our implementation of a prototype of the proposed scheme. In Section V we conclude our proposed scheme and present our future direction along this line.

## II. OVERVIEW

In this section, an overview towards a Software Defined Networking-enabled multicast scheme is proposed which can achieve high bandwidth utilization in cloud computing networks. The goal is to thoroughly employ the route diversity in various network topologies, to achieve ultra-high bandwidth utilization thus improving the performance of multicast communication in cloud networks. The proposed approach has several crucial design decisions that lead to the desired property of the scheme.

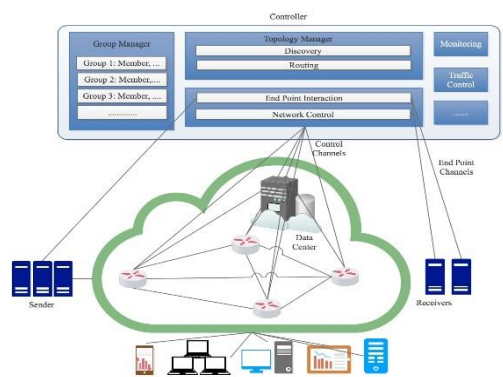

Fig1: Software Defined Networking-Enabled Multicast Scheme for cloud Computing

Figure 1 gives an overview of the Software Defined Networking-enabled multicast scheme framework in cloud networks revealing the various elements, the interactions between the components and the interactive connections between the controller components and the cloud networks components. In this figure, some of the primary components of Software Defined Networking-enabled multicast scheme in cloud network are shown. Various other functionalities can be connected by loading different modules into the controller of SDN, such as traffic control, network status monitoring etc.

The SDN controller needs only reveal an interface to the network. The SDN controller is placed close to the cloud network so that it can control different components of the cloud network. The interface represents the relationship between the network elements (end-points and intermediate nodes) of the cloud networks and the corresponding controller component (network control and end-point interaction). Reasonably, both the end-points and the intermediate nodes of the cloud networks can trigger network modifications, which raises various events in the controller. The interaction component utilizes different events, invoking the corresponding processing logic inside the controller for further processing, and then replies the processing results if needed. Upon getting network events from various components of the cloud network, the controller requires to carry out different jobs. For instance, when a receiver requests to join or leave a group, the group management component needs to validate the request with existing group information, perform the membership operation, and reply the corresponding information; or when a link malfunction is identified, the topology management needs to adjust the topology information, and also, recalculate all existing routes for accurate transmission in cloud networks.

A key component is the routing component. It carries out the task of calculating the routes near the cloud networks regarding a request for transmission. In the proposed scheme, each transmission session is not associated with a particular tree, but various different multicast trees each with varying capacity shares. There are various algorithms to calculate multiple Steiner trees given the topology. It is important to choose the algorithm regarding many constraints for cloud computing, such as computation complexity, resource consumption, etc.

There are diverse benefits in using the SDN control plane model in cloud computing, i.e., using a logically centralized controller of SDN as the control plane near the cloud of the network. The SDN control model is a short-cut in solving many of the intrinsic limitations of multicast communication. Improved scalability for cloud networks. SDN can considerably increase the scalability of multicast for cloud networks, due to the excessive amount of resource the controller can utilize as a server application. In traditional cloud computing, either the switches or the routers have adequate computing and storage resources for multicast route computing and group management [9].

The controller placed near the cloud of the network has much more resources for management and computation compared to traditional cloud networks. A typical PC with 4GB memory space can support thousands to tens of thousands of multicast groups. A further improvement may come from implementing a "logically centralized" controller architecture, where there are multiple identical physical controllers balancing the control load network-wide.

- **Overhead offloading**: SDN offloads most of the computation and storage overhead to the controller, which releases the resources on the forwarding elements. While traditional switches or routers placed near the cloud of networks need to perform various tasks such as routing, group management, and information propagation. Whereas, SDN placed near the cloud of the networks have all their power can be used for forwarding, leading to improvements in the speed of forwarding in cloud computing.
- Optimality of Control: SDN placed near the cloud of the network facilitates the control plane to perform globally optimal control decisions by implementing a global network view near the cloud of the network rather than a limited local view of cloud networks. In traditional cloud networks, the forwarding elements form their own decisions based on the local view (or more precisely, a delayed view) of the network, which is aggregated from only their neighbours in the clouds. The limited scope of view limits the quality of decision made by switches or routers, which further degrades the network performance of traditional cloud networks. On the contrary, the global view of the controller in SDN placed near the cloud of the networks provides the opportunity to achieve global optimality when making local decisions, i.e., optimizing some network objectives globally in routing and forwarding.

Besides the above benefits of SDN in cloud computing that directly improves the performance of multicast communication, SDN can also provide better flexibility in control and programmability for cheaper network management. All these benefits lead us to choose SDN as the control plane solution of our proposed multicast scheme.

### III. OVERALL DESIGN

The proposed idea of Software Defined Networking-Enabled multicast scheme for cloud computing is to construct multiple multicast trees for every multicast session and distribute traffic load on each tree in cloud networks. Following the methodology of MPTCP [10], each multicast tree is signified as a sub-flow in the multicast session. Each sub-flow manages its private transmission state, reliability mechanism, and congestion control. Each sub-flow is unaware of the presence of other sub-flows that belong to the same multicast session, thus maintaining the simplicity of flow management. To accomplish this simplicity while achieving low overhead in transmission and on intermediate nodes, the SDN is introduced at the cloud of the network to carry out heavy-load jobs in the transmission process. Each multicast session is consisting of four components: the sender, the forwarding elements, the receivers and the SDN controller. Each multicast session has exactly one sender, usually more than one receivers and is managed by exactly one SDN controller. The data center present in the cloud computing is also managed by the SDN controller which increase the efficiency to process data at the cloud of the network. In the following subsections, the operations on each component are illustrated respectively.

*A. Sender Operations*

The sender present in the cloud of the network has several responsibilities in the multicast session. First, the sender needs to initiate the transmission of a data block. The initiation process is composed of several procedures. In the first place, the sender needs to communicate with the multicast controller present in the cloud computing. The sender will first send the information about the transmission to the controller by sending out a designated packet towards a specific IP address, namely, the management address. The information includes the multicast group this sender is sending to, the length of the data block, etc. This management address is either pre-configured or not matching in any switch flow table, both will lead to the packet being sent to the controller. Upon retrieval of this initiation packet, the controller will carry out calculations associated with this multicast session, and reply the information necessary for the sender to correctly initiate all the multicast sub-flows and start the transmission. When the sender receives this initiation reply, it will initiate all the sub-flows and begin transmission. In the process of transmitting, the sender must still listen to any traffic coming from the controller, in case that the status changes at the cloud of the network. When the sender receives packet about network dynamics, it will then change the behaviour of transmitting according to the control packet. After the transmission of a traffic block, the sender will send another packet towards the management address, indicating the end of the transmission. No controller reply is needed for the ending packet, in order to save the sender for other tasks.

*B. Intermediate Node Operations*

The intermediate nodes of the cloud computing are only responsible for packet forwarding based on their flow tables. They will basically match on the address field (the session identifier) and the sub-flow identifier field. Since both the group management and the decision making is pre-done by the controller, no computational resource is wasted on these nodes. Thus, the intermediate nodes are expected to achieve their best performance in packet forwarding.

*C. Receiver Operations*

First, a receiver joins a multicast group by sending "Group Join" message to the corresponding management address. Upon retrieval of the join message, the controller updates the management database and replies with the necessary information about the multicast group. After joining the group, the receivers in a multicast session basically plays the role of traffic aggregator and network state monitor. First, the receivers need to aggregate traffic from all multicast sub-flows of the corresponding multicast session into the sequential flow of data, which can be directly consumed by the upper layer. The aggregation process basically follows the MPTCP method, thus will not be discussed in this paper.

The receivers also perform reliability functionalities. There have been many existing reliability mechanisms for multicast transportation [1], [2], [11]. The proposed scheme is free to perform any of the existing reliability protocols, only that all mechanisms are performed in the sub-flow level. Besides, the receivers can also reply receiving statistics to the controller through the management address. This can provide some sort of feedback for the in-time update of the network status. A receiver leaves a group by sending "Group Leave" message to the corresponding management address.

*D. Controller Operations*

The controller present near the cloud of the network takes most of the jobs in maintaining a multicast session. Below are some of the responsibilities of the controller.

- **Group management**: The controller manages the information of all multicast groups and sessions in the cloud network. It processes joining and leaving of group members, as well as periodically checking the status of each member. By offloading the task of group management to the controller, the intermediate nodes in the cloud network can then devote all their efforts to packet forwarding, greatly improving the performance and the scalability of the multicast scheme for cloud computing.
- **Topology service and routing**: The controller of the cloud computing keeps track of the global view of the whole cloud network, and thus is optimal in providing network view and calculating best routes for nearby components present in the cloud. Upon the retrieval of the transmission initiation message from each sender near the cloud, the controller computes the optimal routes regarding of the session and replies the information towards the sender, thus initiating the transmission. If there is not enough capacity in the cloud network, or some receivers are temporarily unreachable, the controller will either pause or drop the initiation message.
- **Flow table updating:** Prior to initiating each session of transmission, the controller will first configure each intermediate nodes of the cloud network in the routes of the session, updating their flow tables in order to provide the correct routing functionality. Also, when the network is undergoing dynamics, the controller need to recompute all the routes and update the corresponding switch tables.

Besides these basic functionalities, the controller can apply various policies, such as middlebox enforcement etc. In this paper, the primary functionalities required by a multicast scheme are examined.

*E. Management Communication*

It is notable to mention how the end-hosts at the cloud of the network communicate with the SDN controller. Basically, the communication between end-hosts and the controller has three types: receiver-initiated, sender-initiated and controller-initiated. The corresponding message formats are defined in the management protocol.

Receiver-initiated communication basically includes two kinds of messages: request for group join, and notification of group leave. When the receiver wishes to join a group, it sends the group join message to the controller, and wait for the reply from the controller about joining status; when it wishes to leave a group, it sends the group leave message and quit without reply. Similarly, sender-initiated communication basically involves session initialization and session end, the former with a reply and the latter without. The controller-initiated messages mostly involve network or group status updates.

There are two choices for the communication between end hosts and the controller: connectional and connectionless. In this proposed scheme, the connectionless communication for information exchange between end-hosts and the controller is considered. This is due to three reasons:

1) Connectional communication will incur extra communication overhead, which will influence the response time off from the controller;
2) Direct connectional communication between end hosts and the controller may raise security issues;
3) The communication is applied in a simple request-reply manner, in which case the requester can keep the full responsibility of retransmitting the request or verifying the transmitted reply, thus the benefits of connectional communication is minor in this case.

The communication takes place in the following process. For end-host-initiated communication, the requester manipulates the request message based on the protocol sends the message towards a specific virtual management address and listens to the send port for the reply. The controller receives all packets sent to this management address parses the requests according to the protocol, verifies and processes the requests, manipulates the reply message, and sends to the source addresses and ports based on request packets. For controller-initiated communication, the controller simply manipulates the update messages and sends to corresponding end-hosts based on its group and session data stored.

A crucial design in this process is the virtual management address. This address is actually not bound to an agent in the network, reserved for the use of management only. Specifically, this address is defined in the management protocol, which is known to both the controller and the end-hosts. In establishing the network, the controller will set up a flow entry in each switch, matching all packets towards this address with the action of "forwarding to the controller", prior to any communication behaviour in the network. In using this virtual address rather than the physical address of the controller, it can keep its identity and location private to only the network administrator, thus securing the controller from anonymous connections.

*F. Routing*

The proposed scheme achieves high bandwidth utilization through transmission over multiple trees for each session. It is important to understand how the cloud network routes the packets belonging to one session but multiple trees. In transmission, the sender needs to split the data packets overall the trees it can utilize, and the intermediate nodes need to distinguish data packets on different routes belonging in one session and perform different forwarding decisions. This is done by encoding the route identifier of each route into the packet header.

When initializing the transmission session, the sender and the controller will exchange information regarding the set-up of the session, which includes the number of trees in this session, as well as the route identifier of each tree. On the sender side, this route identifier is thus encoded into the packet header of each data packet, based on the load splitting policy installed. On the controller side, when configuring the network for this session, it needs to set up flow entries for each tree on corresponding switches, matching the field where the route identifier is encoded, and performing the corresponding forwarding behaviour. This encoding should be pre-defined in the proposed scheme, which is known to both the sender and the controller. Applicable fields include VLAN ID or MPLS tag.

### IV. CASE STUDY

Our performance evaluation is conducted on GENI test-bed [12], a well-known network emulator with special supports for OpenFlow simulation which uses POX as a controller. In our simulation, we first build a designated GENI test-bed topology for testing. After that, we run our controller, receiver and sender program on corresponding virtual nodes in the topology. We count the data amount received at each receiver program as our main metrics of performance evaluation. Our GENI topology is shown in Figure 2. We have one sender *s*, and three receivers *r1*; *r2*; *r3*.

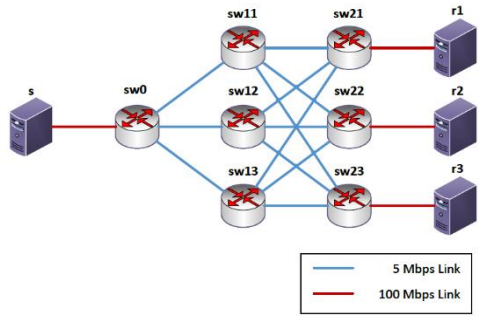

Fig 2: GENI Topology for cloud Computing

To validate our multitree scheme, we set cloud links near the cloud of the network to have a capacity of $100Mbps$, while core links have a capacity of $5Mbps$. This is a common scenario, as for example in data center networks, the core links are usually the bottleneck of the whole of the cloud computing. The traffic is generated on the sender, where fixed-length meaningless bytes are filled in the sender buffer and sent on the different routes.

The proposed scheme is compared to a singletree multicast scheme in some experiments. Without loss of generality, we implement the single-tree routing algorithm using BFS as well. This is because, in this paper, we don't take into consideration QoS metrics other than bandwidth such as latency or packet loss. Therefore, in a homogeneous network topology such as the one we used, the bandwidth property of any single-tree routing scheme should be the same.

Figure 3 shows the application-level goodput from the view of the three receivers, using both a single tree scheme and the proposed scheme. It is clear that the goodput received through the proposed scheme is much higher than that through a simple single-tree multicast scheme. Also, the average goodput of scheme is nearly 3 times of the goodput of single-tree multicast.

This is because in our test topology shown in Figure 2, there are three cloud-disjoint trees from *s* to *r1*; *r2*; *r3*, each goes through *sw11*; *sw12*; *sw13* respectively. By fully utilizing the path diversity between *sw11*; *sw12*; *sw13* and *sw21*; *sw22*; *sw23*, the proposed scheme can achieve three times the bandwidth utilization of a traditional single-tree scheme.

One possible concern about Figure 3 may be the saw-toothlike vibration of the traffic. This is actually because we implemented our transmission upon the UDP protocol, who lacks the basic congestion control mechanisms as TCP has. Specifically, this sawtooth vibration is the consequence of intermediate nodes unfairly drops the incoming packets in a simple drop-tail manner. Therefore, either a well-designed congestion control mechanism or adding fair-queuing functionality onto intermediate nodes shall alleviate the problem.

Similar to Figure 3, Figure 4 shows the time for transmission versus the number of data bytes transmitted. The data is collected by explicitly specifying in the controller up to how many trees one transmission session can use, thus the results for a number of trees equal to *1*; *2*; *3* respectively. The results confirm our demonstration in Figure 3. With an increased number of trees being used for transmission, the transmission time of the same amount of data can be greatly reduced. Note that in the topology, all the three trees are cloud-disjoint, in which case, multi-tree multicast can obtain the best gain.

When multiple trees are sharing a bottleneck link, there may be degradation in performance. But it is clear that even in the worst-case scenario, where all multiple trees share a single link, the performance of proposed scheme will be at least as good as any single-tree multicast scheme, given the same algorithm in finding the multicast tree.

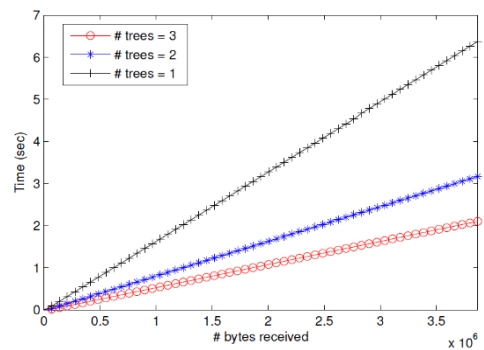

Fig 4: Transmission Time vs. Bytes Received

While Figure 3 and Figure 4 give a general view of the performance of proposed scheme, in Figure 5, we show how the load is balanced among the trees. We take the application-level goodput of receiver *r1* and plot the goodput bandwidth received on each of the three multicast trees.

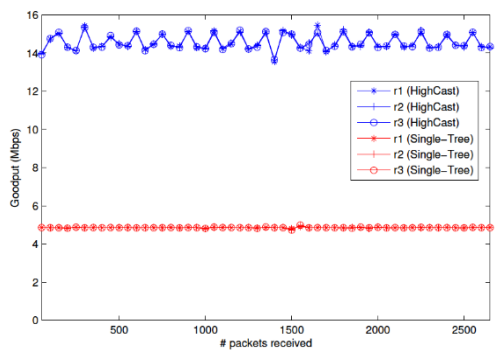

Fig 3: Application-level Goodput of Receivers

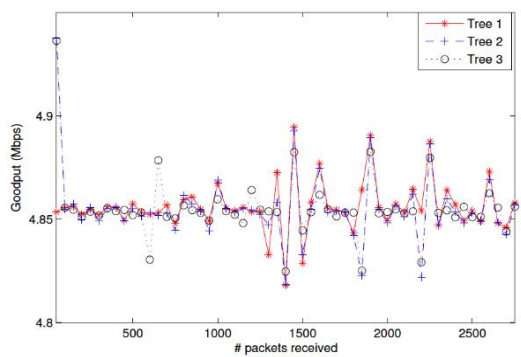

Fig 5: Application-level Goodput of Trees on Receiver *r1*

From Figure 5, although there seems some difference in the throughput among the three trees, in most cases they show the same traffic pattern. As above, the vibration is due to the drop-tail the behaviour of the switches. From this figure, we can further figure out that the vibration is mainly caused by the drop-tail the behavior of the output queues at switch *sw0*.

The idea here is that 1) switch *sw0* is the first and possibly the only bottleneck that every sub-flow encounters in the network, and 2) traffic vibration on different trees follows a very similar pattern. It is not likely that every sub-flow follows the same traffic pattern when the bottleneck is not shared among all sub-flows.

The above simulation results show that our proposed scheme can achieve ultra-high bandwidth utilization compared to traditional single-tree multicast schemes. Also, by using designated load balancing and traffic engineering mechanisms, we can achieve balanced load among all trees.

## V. CONCLUSION

In this paper, we studied the existing limitations and challenges of various conventional multicast communication schemes in cloud computing, as well as how SDN can solve the problems and enable an effective and efficient multicast scheme design for cloud computing. The proposed scheme, an SDN-enabled multicast scheme for cloud computing that, not only solves the existing challenges and limitations by introducing the flexible and powerful control plane of SDN, but also achieves ultra-high bandwidth utilization through fully utilizing the path diversity in the given cloud network topology.

The proposed scheme for cloud computing is both effective, i.e., it achieves the basic requirements of a multicast communication scheme, and efficient, i.e., it shows significant performance improvement compared to existing solutions. Besides that, the proposed scheme is also flexible and extensible. With the help of SDN, one can easily load different policies and algorithms into the existing framework, such as different multi-tree routing algorithms, more complicated traffic control mechanisms, etc. The proposed scheme is validated by actually implementing a prototype of cloud computing, and evaluating the scheme using different metrics. Experiment results confirm the proposed argument is both effective and efficient. While there are still a lot of future directions along this line, the proposed scheme shows promising properties in improving network performance even with its primary architecture for cloud computing.